\documentclass[10pt,twocolumn,prl,aps,amssymb,amsmath,showpacs,tightenlines]
{revtex4}

\newcommand{\half}{\mbox{$\textstyle \frac{1}{2}$}}

\begin{document}

\title{Semiclassical analysis of a complex quartic Hamiltonian}

\author{Carl~M.~Bender${}^*$, Dorje~C.~Brody${}^\dagger$, and
Hugh~F.~Jones${}^\dagger$}

\affiliation{${}^*$Department of Physics, Washington University,
St. Louis, MO 63130, USA \\ ${}^\dagger$Blackett Laboratory,
Imperial College, London SW7 2BZ, UK}

\date{\today}

\begin{abstract}
It is necessary to calculate the $\cal C$ operator for the
non-Hermitian $\cal P T$-symmetric Hamiltonian $H=\half
p^2+\half\mu^2x^2-\lambda x^4$ in order to demonstrate that $H$
defines a consistent unitary theory of quantum mechanics. However,
the $\cal C$ operator cannot be obtained by using perturbative
methods. Including a small imaginary cubic term gives the
Hamiltonian $H=\half p^2+\half \mu^2x^2+igx^3-\lambda x^4$, whose
$\cal C$ operator {\it can} be obtained perturbatively. In the
semiclassical limit all terms in the perturbation series can be
calculated in closed form and the perturbation series can be
summed exactly. The result is a closed-form expression for $\cal
C$ having a nontrivial dependence on the dynamical variables $x$
and $p$ and on the parameter $\lambda$.
\end{abstract}


\maketitle

In this paper we consider a quantum system described by the
Hamiltonian
\begin{eqnarray}
H=\half p^2+\half\mu^2 x^2+igx^3-\lambda x^4, \label{e1}
\end{eqnarray}
where $g$ is real and nonzero and $\lambda\geq0$. Note that the
potential is complex and that when $\lambda$ is positive the
potential is unbounded below. This Hamiltonian is not Hermitian in
the conventional sense, where Hermitian conjugation is defined as
combined transpose and complex conjugate. Nevertheless, the
eigenvalues $E_n$ are all real, positive, and discrete. This is
because $H$ possesses an unbroken $\cal PT$ symmetry \cite{B1,B2},
which means that $H$ and its eigenstates $\psi_n(x)$ are invariant
under space-time reflection. Here, $\cal P$ denotes the spatial
reflection $p\to-p$ and $x\to-x$, and $\cal T$ denotes the time
reversal $p\to-p$, $x\to x$, and $i\to-i$.

Many $\cal PT$-symmetric quantum-mechanical Hamiltonians have been
studied in the recent literature \cite{R1,R2,R3,R4}. However, the
Hamiltonian (\ref{e1}) is special because when $g\neq0$ the
boundary conditions on the eigenfunctions may be imposed on the
real-$x$ axis, as opposed to the interior of a wedge in the
complex-$x$ plane, as we will now show: The quantization condition
satisfied by the eigenfunctions requires that $\psi_n(x)$ must
vanish exponentially in a pair of wedges in the complex-$x$ plane.
These wedges are symmetrically placed with respect to the
imaginary axis. The wedges have an angular opening of $60^\circ$
and lie below the positive and negative real-$x$ axes with the
upper edges of the wedges lying on the real axis. Using a WKB
approximation we can determine the asymptotic behavior of the
eigenfunctions, and we find that the exponential decay of these
wave functions is given by
\begin{equation}
\psi_n(x)\sim
e^{\pm\sqrt{2\lambda}[ix^3/3+gx^2/(4\lambda)]}\quad(|x|\to\infty).
\label{e2}
\end{equation}
Thus, the wave functions are oscillatory on the positive and
negative real-$x$ axes when $g=0$. However, when $g$ is nonzero
the wave functions decay exponentially on the real axis as well as
in the interiors of the wedges. Thus, taking $g\neq0$ allows us to
treat $x$ as a real variable and to perform calculations on the
real axis.

Being able to treat $x$ as real is crucial. The domain of the
eigenfunctions of $H=\half p^2+\half \mu^2x^2-\lambda x^4$ is the
interior of a pair of $60^\circ$-wedges in the lower-half
$x$-plane. Under space reflection $x\to-x$, this domain changes to
the interior of a pair of $60^\circ$-wedges in the {\it
upper}-half plane. Therefore, this Hamiltonian is not parity
symmetric. However, when $g\neq0$, the domain of the
eigenfunctions of $H$ in (\ref{e1}) includes the real-$x$ axis.
Thus, on the real-$x$ axis, the parity operator $\cal P$ commutes
with the $-x^4$ operator. This fact enables us to perform in this
paper a perturbative calculation of $\cal C$. The $\cal C$
operator is needed to formulate a consistent quantum theory
described by the non-Hermitian Hamiltonian (\ref{e1}).

To make sense of $H$ in (\ref{e1}) we must identify the Hilbert
space spanned by the eigenfunctions of $H$ and then construct for
this space an inner product that is positive definite. As shown in
Ref.~\cite{B2}, an inner product defined with respect to $\cal
PT$-conjugation leads to an indefinite metric of the type
investigated earlier by Lee and Wick~\cite{LW}. However, an inner
product defined with respect to $\cal CPT$-conjugation leads to a
positive definite metric, and hence positive probabilities
\cite{B2,M}. Here $\cal C$ denotes a linear operator analogous to
the charge operator in particle physics. The operator $\cal C$
commutes with the Hamiltonian and its square is unity, so its
eigenvalues are $\pm1$. Because $\cal C$ commutes with $H$, the
time evolution of the theory is unitary; that is, the norm of a
vector is preserved in time. Given the operator $\cal C$, we can
construct the {\it positive} operator $e^Q= \cal{CP}$, which can,
in turn, be used to construct by a similarity transformation an
equivalent Hamiltonian ${\tilde H}\equiv e^{-Q/2}He^{Q/2}$. The
Hamiltonian $\tilde H$ is Hermitian in the conventional sense
\cite{M}, but it is a nonlocal function of the operators $x$ and
$p$.

Thus, the key step in formulating a consistent quantum theory
based on the Hamiltonian (\ref{e1}) is to calculate the operator
$\cal C$. When $\lambda=0$, one can use perturbation theory to
calculate the $\cal C$ operator as a series in powers of $g$
\cite{BMW}. However, for the more interesting case of a negative
quartic interaction ($g=0$, $\lambda>0$), a perturbative
calculation of $\cal C$ using conventional Poincar\'e asymptotics
fails because to all orders in powers of $\lambda$ the operator
$Q$ vanishes. Only nonperturbative techniques such as
hyperasymptotics (asymptotics beyond all orders) \cite{BERRY} can
be used to find the $\cal C$ operator for the Hamiltonian $H=\half
p^2+\half \mu^2x^2-\lambda x^4$.

The analysis in this paper is based on the observation that when
$g\neq0$, no matter how small, it is possible to use perturbative
methods to calculate $\cal C$. Our perturbative calculation is
organized as follows: First, we introduce the small positive
parameter $\epsilon$ into the Hamiltonian (\ref{e1}) and consider
\begin{eqnarray}
H=\half p^2+\half\mu^2x^2+i\epsilon gx^3-\epsilon^2\lambda x^4.
\label{e4}
\end{eqnarray}
We seek a perturbation series in powers of $\epsilon$. The
coefficient of $\epsilon^n$ in this perturbation series is
complicated, and thus our second step is to simplify the
coefficient by making a semiclassical approximation in which we
only retain leading order terms in Planck's constant $\hbar$. The
result is a series in powers of $g$, and since we may take $g$
arbitrarily small, our third step is to simplify the coefficient
further by omitting all contributions from higher powers of $g$.
The resulting infinite series can then be summed exactly and in
closed form. Once the summation is performed, our fourth step is
to set $\epsilon=1$ to obtain the semiclassical approximation to
$\cal C$ for the Hamiltonian (\ref{e1}). The Hamiltonian
(\ref{e4}) was first considered by Banerjee \cite{BAN}. In
Ref.~\cite{BAN} the first seven terms in the perturbation
expansion for the $\cal C$ operator are calculated (but not in the
semiclassical regime).

We begin our analysis by recalling that the $\cal C$ operator can
be expressed in the form ${\cal C}=e^Q{\cal P}$, where $Q=Q(x,p)$
is a function of the quantum dynamical variables $x$ and $p$. In
earlier work we showed that $\cal C$ can be determined by
searching for an operator that satisfies the following three
conditions \cite{B3}:
\begin{eqnarray}
({\rm i})~[{\cal C},{\cal PT}]=0,\quad({\rm ii})~{\cal C}^2={\bf
1},\quad({\rm iii})~[H,{\cal C}]=0. \label{e5}
\end{eqnarray}
Substituting ${\cal C}=e^Q{\cal P}$ into (i), we obtain the
condition $Q(x,p)=Q( -x,p)$, so $Q(x,p)$ is an even function of
$x$. Substituting ${\cal C}=e^Q{\cal P}$ into (ii), we get
$Q(x,p)=-Q(-x,-p)$. Since $Q(x,p)$ is even in $x$, it is odd in
$p$. Finally, condition (iii) reads
\begin{eqnarray}
[H,e^Q{\cal P}]=0. \label{e6}
\end{eqnarray}

Our objective is to determine the expression for the operator
$Q(x,p)$, when $H$ is given by (\ref{e4}). Let us write the
Hamiltonian (\ref{e4}) in the form
\begin{eqnarray}
H=H_0+\epsilon H_1+\epsilon^2H_2, \label{e7}
\end{eqnarray}
where $H_0$ is the Harmonic oscillator Hamiltonian, $H_1=igx^3$,
and $H_2=- \lambda x^4$. The commutation relation (\ref{e6}) then
implies
\begin{eqnarray}
H_0e^Q{\cal P}\!&-&\!e^Q{\cal P}H_0+\epsilon\left(H_1e^Q{\cal
P}-e^Q{\cal P}H_1
\right)\nonumber\\
&+&\!\epsilon^2\left(H_2e^Q{\cal P}-e^Q{\cal P}H_2\right)=0.
\label{e8}
\end{eqnarray}
Under parity we have
\begin{eqnarray}
\!\!{\cal P}H_0{\cal P}=H_0,\!\!\!\quad{\cal P}H_1{\cal
P}=-H_1,\!\!\!\quad{\rm and}\!\!\!\quad{\cal P}H_2{\cal P}&=&H_2.
\label{e9}
\end{eqnarray}
As noted above, $H_2$ for $x$ real commutes with $\cal P$ because
$g$ is nonzero.

Substituting these relations into (\ref{e8}) and multiplying $\cal
P$ from the right, we obtain
\begin{eqnarray}
e^QH_0-H_0e^Q&=&\epsilon\left(e^QH_1+H_1e^Q\right)\nonumber\\
&&-\epsilon^2\left(e^QH_2-H_2e^Q\right). \label{e10}
\end{eqnarray}
We then multiply by $e^{-Q}$ on the left and get
\begin{eqnarray}
H_0-e^{-Q}H_0e^Q&=&\epsilon\left(H_1+e^{-Q}H_1e^Q\right)
\nonumber\\ &&-\epsilon^2\left(H_2-e^{-Q}H_2e^Q\right).
\label{e11}
\end{eqnarray}

In order to analyze (\ref{e11}) we make use of the
Campbell-Baker-Hausdorff relation
\begin{eqnarray}
e^{-Q}He^Q&=&H+[H,Q]+{\textstyle\frac{1}{2!}}[[H,Q],Q] \nonumber\\
&&+{\textstyle\frac{1}{3!}}[[[H,Q],Q],Q]+\cdots \label{e12}
\end{eqnarray}
and the fact that $Q$ can be expanded as a power series in
$\epsilon$:
\begin{eqnarray}
Q=\epsilon Q_1+\epsilon^3Q_3+\epsilon^5Q_5+\cdots. \label{e13}
\end{eqnarray}
Substitution of (\ref{e13}) into the right side of (\ref{e12})
yields
\begin{eqnarray}
e^{-Q}He^Q&=&H+\epsilon[H,Q_1]+{\textstyle\frac{1}{2!}}
\epsilon^2[[H,Q],Q]
\nonumber\\
&&\hspace{-1.0cm}+\epsilon^3\left([H,Q_3]+
{\textstyle\frac{1}{3!}}[[[H,Q_1],
Q_1] ,Q_1]\right)\nonumber\\
&&\hspace{-1.0cm}+\epsilon^4\left({\textstyle\frac{1}{2!}}
[[H,Q_3],Q_1]+
{\textstyle\frac{1}{2!}}[[H,Q_1],Q_3]\right.\nonumber\\
&&\hspace{-1.0cm}\quad\left.+{\textstyle\frac{1}{4!}}
[[[[H,Q_1],Q_1],Q_1],Q_ 1] \right)\nonumber\\
&&\hspace{-1.0cm}+\epsilon^5\left({\textstyle\frac{1}{5!}}
[[[[[H,Q_1],Q_1],Q _1],
Q_1],Q_1]\right.\nonumber\\
&&\hspace{-1.0cm}\quad+{\textstyle\frac{1}{3!}}
[[[H,Q_1],Q_1],Q_3]+
{\textstyle\frac{1}{3!}}[[[H,Q_1],Q_3],Q_1]\nonumber\\
&&\hspace{-1.0cm}\quad\left.+{\textstyle\frac{1}{3!}}
[[[H,Q_3],Q_1],Q_1]+[H, Q_5] \right)+\cdots. \label{e14}
\end{eqnarray}
Inserting the expansion (\ref{e14}) into (\ref{e11}) and equating
coefficients of powers of $\epsilon$, we obtain the following set
of identities:
\begin{eqnarray}
&&[Q_1,H_0]=2H_1,\nonumber\\
&&[Q_3,H_0]={\textstyle\frac{1}{3!}}[[[H_0,Q_1],Q_1],Q_1]+
{\textstyle\frac{1}{2!}}[[H_1,Q_1],Q_1]\nonumber\\
&&\hspace{0.2cm}+[H_2,Q_1],\nonumber\\
&&[Q_5,H_0]={\textstyle\frac{1}{5!}}[[[[[H_0,Q_1],Q_1],Q_1]
,Q_1],Q_1]\nonumber\\
&&\hspace{0.2cm}+{\textstyle\frac{1}{4!}}[[[[H_1,Q_1],Q_1],Q_1]
,Q_1]\nonumber\\
&&\hspace{0.2cm}+{\textstyle\frac{1}{3!}}[[[H_0,Q_1],Q_1],Q_3]+
{\textstyle\frac{1}{3!}}[[[H_0,Q_1],Q_3],Q_1]\nonumber\\
&&\hspace{0.2cm}+{\textstyle\frac{1}{3!}}[[[H_0,Q_3],Q_1],Q_1]+
{\textstyle\frac{1}{3!}}[[[H_3,Q_1],Q_1],Q_1]\nonumber\\
&&\hspace{0.2cm}+{\textstyle\frac{1}{2!}}[[H_1,Q_3],Q_1]+
{\textstyle\frac{1}{2!}}[[H_1,Q_1],Q_3]+[H_2,Q_3], \label{e15}
\end{eqnarray}
and so on. These identities correspond to the coefficients of
$\epsilon$, $\epsilon^3$, and $\epsilon^5$. The coefficients of
the even powers of $\epsilon$ are redundant because they can be
derived from the coefficients of the lower odd powers of
$\epsilon$. For example, the coefficient of $\epsilon^2$ gives
$[[H_0,Q_1],Q_1]=-2[H_1,Q_1]$, which follows from the first
relation in (\ref{e15}).

We now perform a semiclassical approximation in which we only
retain terms to leading order in $\hbar$. That is, we use the fact
that each operator $Q_i$ in (\ref{e13}) has a semiclassical
expansion of the form
\begin{eqnarray}
Q_i=\frac{1}{\hbar}Q_i^{(-1)}+Q_i^{(0)}+\hbar
Q_i^{(1)}+\hbar^2Q_i^{(2)}+\cdots,
\end{eqnarray}
and discard all but the leading terms $Q_i^{(-1)}$ for
$i=1,3,5,\ldots$. Because we consider only the leading terms
$Q_i^{(-1)}$, in what follows we omit the superscript and write
$Q_i$ for simplicity of notation.

We remark that in a semiclassical approximation, once a
commutation relation is performed, we can regard $x$ and $p$ as
classical variables and hence issues relating to operator ordering
need not be considered. In this connection, the following relation
applicable in semiclassical approximation is useful:
\begin{eqnarray}
[p^ax^b,p^cx^d]=i\hbar(bc-ad)p^{a+c-1}x^{b+d-1}. \label{e16}
\end{eqnarray}
This is a special case of the Poisson bracket relation
\begin{equation}
\{F(x,p),G(x,p)\}=i\left(\frac{\partial F}{\partial
x}\frac{\partial G}{\partial p}-\frac{\partial F}{\partial
x}\frac{\partial G}{\partial p}\right). \label{e16.5}
\end{equation}

Using the semiclassical commutation relation (\ref{e16}), we solve
the first equation in (\ref{e15}) for $Q_1$ and obtain
\begin{eqnarray}
Q_1=-\frac{4g}{\mu^4\hbar}\Big[{\textstyle\frac{1}{3}}
p^3+\half\mu^2px^2\Big ]. \label{e17}
\end{eqnarray}
Substituting (\ref{e17}) into the second relation of (\ref{e15})
allows us to determine $Q_3$ as
\begin{eqnarray}
Q_3&=&-\frac{4^2\lambda
g}{\mu^8\hbar}\Big[{\textstyle\frac{2}{5}}p^5+\mu^2p^3
x^2+\half\mu^4px^4\Big]\nonumber\\
&&+\frac{4^2g^3}{\mu^{10}\hbar}\Big[ {\textstyle\frac{8}{15}}p^5
+{\textstyle\frac{5}{6}}\mu^2p^3x^2+\half\mu^4px^4\Big].
\label{e18}
\end{eqnarray}
Similarly, substituting (\ref{e17}) and (\ref{e18}) into
(\ref{e15}), we deduce that
\begin{eqnarray}
Q_5&=&-\frac{4^3\lambda^2g}{\mu^{12}\hbar}\Big[
{\textstyle\frac{4}{7}}p^7+2\ mu^2
p^5x^2+2\mu^4p^3x^4+\half\mu^6px^6\Big]\nonumber\\
&&\hspace{-0.4cm}+\frac{4^3\lambda
g^3}{\mu^{14}\hbar}\Big[{\textstyle\frac{16}
{7}}p^7+6\mu^2p^5x^2+{\textstyle\frac{16}{3}}
\mu^4p^3x^4+{\textstyle\frac{7} {4}}
\mu^6px^6\Big]\nonumber\\
&&\hspace{-0.4cm}-\frac{4^3g^5}{\mu^{16}\hbar}\Big[
{\textstyle\frac{5}{3}}p^ 7+
{\textstyle\frac{17}{6}}\mu^2p^5x^2+{\textstyle
\frac{8}{3}}\mu^4p^3x^4+\mu^6 p x^6\Big]. \label{e19}
\end{eqnarray}

Continuing in this manner, we can determine the perturbative
expansion of $Q$ explicitly. Observe, however, that $Q_{2n+1}$ for
each $n=0,1,2,\ldots$ is an odd polynomial of $g$ of degree
$2n+1$. This follows from (\ref{e15}) if we notice that $H_1$ and
hence $Q_1$ are proportional to $g$ whereas $H_0$ and $H_2$ are
independent of $g$. Because we assume that the Hamiltonian
(\ref{e1}) has a weak cubic interaction, the value of the coupling
$g$ is small. Therefore, we may omit terms of order $g^3$ and
higher from the expansion of $Q$. To first order in $g$ the set of
identities in (\ref{e15}) reduces to the following simpler set of
relations:
\begin{eqnarray}
\begin{array}{ccl}\vspace{0.1cm}\big[H_0,Q_1\big]
&=&-2H_1\\
\vspace{0.1cm}\big[H_0,Q_3\big]
&=&\big[Q_1,H_2\big]\\
\vspace{0.1cm}\big[H_0,Q_5\big]
&=&\big[Q_3,H_2\big]\\
\vspace{0.1cm}\big[H_0,Q_7\big] &=&\big[Q_5,H_2\big]\\ \vdots &&
\label{e20}
\end{array}
\end{eqnarray}

From these relations we deduce that to first order in $g$, $Q_7$
is given by
\begin{eqnarray}
Q_7&=&-\frac{4^4\lambda^3g}{\mu^{16}\hbar}\Big[
{\textstyle\frac{8}{9}}p^9+4\ mu^2
p^7x^2+6\mu^4p^5x^4\nonumber\\
&&+{\textstyle\frac{10}{3}}\mu^6p^3x^6+\half\mu^8px^8\Big],
\end{eqnarray}
and that to first order in $g$, $Q_9$ is given by
\begin{eqnarray}
Q_9&=&-\frac{4^5\lambda^4g}{\mu^{20}\hbar}\Big[
{\textstyle\frac{16}{11}}p^{1 1}
+8\mu^2 p^9 x^2+16\mu^4p^7x^4\nonumber\\
&&+14\mu^6p^5x^6+5\mu^8p^3x^8+\half\mu^{10}px^{10}\Big].
\end{eqnarray}
By repeating this procedure and determining $Q_{2n+1}$ for
$n=0,1,2,\ldots$, we deduce, in general, that
\begin{eqnarray}
Q_{2n+1}&=&-g\frac{2^{3n+2}\lambda^n}{\mu^{4n+4}\hbar}p\nonumber\\
&&\hspace{-1.2cm}\times\sum_{k=0}^{n+1}\frac{\mu^{2k}(2n-k+2)!}{2^k
k!(2n-2k+3)! }x^{2k}p^{2n-2k+2}. \label{e23}
\end{eqnarray}

To determine the semiclassical expression for $Q$, we must sum the
product $\epsilon^{2n+1} Q_{2n+1}$ in $n$. For convenience we
define the variables
\begin{eqnarray}
\alpha=\frac{\mu^2 x^2}{2p^2}\quad{\rm
and}\quad\beta=\frac{8\epsilon^2\lambda p^2}{\mu^4}, \label{e24}
\end{eqnarray}
and write
\begin{eqnarray}
\epsilon^{2n+1} Q_{2n+1}=-\frac{4\epsilon gp^3}{\mu^4\hbar}
\beta^n\!\sum_{k=0}^{n+1}\frac{(2n-k+2)!}{k!(2n-2k+3)!}\alpha^k.
\label{e25}
\end{eqnarray}
By summing (\ref{e25}) in $n$ and interchanging the orders of
summation, we can express $Q$ as
\begin{eqnarray}
Q &=& - \frac{4\epsilon gp^3}{\mu^4\hbar}\alpha \sum_{k=0}^\infty
\frac{(\alpha\beta)^k}{(k+1)!} \sum_{n=0}^\infty
\frac{(2n+k+1)!}{(2n+1)!}\beta^n \nonumber \\ && - \frac{4\epsilon
gp^3}{\mu^4\hbar} \sum_{n=0}^\infty \frac{1}{2n+3}\beta^n .
\label{e26}
\end{eqnarray}

To determine the first sum in the right hand side of (\ref{e26}),
we use the identity $(2n+k+1)!=\int_0^\infty
dt\,t^{2n+k+1}e^{-t}$. After performing the resulting integral, we
get
\begin{eqnarray}
&&\alpha\sum_{k=0}^\infty \frac{(\alpha\beta)^k}{(k+1)!}
\sum_{n=0}^\infty \frac{(2n+k+1)!}{(2n+1)!}\beta^n \nonumber \\
&& \hspace{0.5cm}=\frac{1}{2\beta^{\frac{3}{2}}}\left[ \ln\frac{
1-\alpha\beta+\sqrt{\beta}}{1-\alpha\beta-\sqrt{\beta}}-\ln\frac{
1+\sqrt{\beta}}{1-\sqrt{\beta}}\right]. \label{e27}
\end{eqnarray}
The summation on the left side of (\ref{e27}) converges to the
right side provided that the inequality
$\alpha\beta+\sqrt{\beta}<1$ is satisfied. More explicitly, this
inequality reads
\begin{eqnarray}
p<-\epsilon\sqrt{2\lambda}\,x^2+\frac{\mu^2}{2
\epsilon\sqrt{2\lambda}}. \label{e28}
\end{eqnarray}
For $\epsilon\ll1$, the summation on the left side of (\ref{e27})
converges essentially in the entirety of the semiclassical phase
space. An analogous conclusion follows in the limit $\lambda\to0$.
For finite $\epsilon$ and $\lambda$, there is a parabolic region
in the semiclassical phase space in which the operator $Q$
converges. We believe that this region might be associated with
the region in which the corresponding classical trajectories are
confined, although we have not studied this question.

The second term on the right side of (\ref{e26}) gives
\begin{eqnarray}
\sum_{n=0}^\infty\frac{\beta^n}{2n+3}=\frac{1}{
\beta^{\frac{3}{2}}}\left( \frac{1}{2}\ln\frac{1+\sqrt{\beta}}{1-
\sqrt{\beta}}-\sqrt{\beta}\right). \label{e29}
\end{eqnarray}
Note that the left side of (\ref{e29}) converges for $\beta<1$,
which holds automatically if (\ref{e28}) is satisfied.

Combining (\ref{e27}) and (\ref{e29}) and substituting
(\ref{e24}), we finally deduce that to leading order in $g$ the
semiclassical expression for the operator $Q$ associated with the
Hamiltonian (\ref{e4}) is
\begin{eqnarray}
Q&=&-\frac{4\epsilon gp^3}{\mu^4\hbar} \left(\frac{1}{2
\beta^{\frac{3}{2}}}\ln \frac{1-\alpha\beta+\sqrt{\beta}}
{1-\alpha\beta-\sqrt{\beta}}-\frac{1}{\beta} \right)\nonumber \\
&& \hspace{-1.2cm}=\frac{g\mu^2 \sqrt{2}}{16\epsilon^2
\lambda^{\frac{3}{2}}\hbar}\ln \frac{\mu^2-4\lambda
\epsilon^2x^2-2\epsilon\sqrt{2 \lambda}\,p}{\mu^2-4\lambda
\epsilon^2x^2+2\epsilon\sqrt{2\lambda}\,p}
-\frac{gp}{2\epsilon\lambda\hbar}. \label{e30}
\end{eqnarray}
We have performed the summation explicitly, so we may set
$\epsilon=1$ in (\ref{e30}) to obtain the corresponding result for
the Hamiltonian (\ref{e1}). This achieves our objective of finding
the semiclassical approximation to the $\cal C$ operator for this
Hamiltonian. Note that if we expand the right side of (\ref{e30})
for small $\lambda$ and then take the limit $\lambda\to0$, we
recover (\ref{e17}). This is because, to first order in $g$, $Q_1$
is the only term that is not proportional to $\lambda$.

A complete analysis of the $\cal C$ operator for a $-\lambda x^4$
quantum-mechanical theory would be of immense importance because
it could lead to an understanding of its $-\lambda\phi^4$
field-theoretic counterpart. This field theory is asymptotically
free \cite{CCC,AAA,BBB} and might well describe the Higgs sector
in the standard model. Of course, the perturbative method used
here does not apply directly to a pure quartic $-\lambda x^4$
theory, which is inherently nonperturbative; that is, we cannot
set $g=0$ to obtain the semiclassical expression for $Q$ in the
$-\lambda x^4$ theory. However, the work we have presented here is
a first step towards our goal of obtaining a complete
semiclassical and nonperturbative treatment of the $-\lambda x^4$
theory in quantum mechanics.

\vskip1pc CMB thanks the U.S.~Department of Energy and DCB thanks
The Royal Society for support.


\begin{enumerate}

\bibitem{B1} C.~M.~Bender and S.~Boettcher, Phys.~Rev.~Lett.
\textbf{80}, 5243 (1998).

\bibitem{B2} C.~M.~Bender, D.~C.~Brody, and H.~F.~Jones, Phys.
Rev.~Lett. \textbf{89}, 270402 (2002); Am.~J.~Phys. \textbf{71},
1095 (2003).

\bibitem{R1} P.~Dorey, C.~Dunning, and R.~Tateo, J.~Phys.~A:
Math.~Gen. {\bf 34}, 5679 (2001).

\bibitem{R2} F.~Kleefeld, preprint hep-th/0408028 (2004).

\bibitem{R3} M.~Znojil, J.~Math.~Phys. {\bf 46}, 062109 (2005).

\bibitem{R4} For further references see Proceedings of the First,
Second, and Third International Workshops on Pseudo-Hermitian
Hamiltonians in Quantum Mechanics (ed.~by M.~Znojil) in
Czech.~J.~Phys. {\bf 54}, issues \#1 and \#10 (2004) and {\bf 55}
(2005) (to appear).

\bibitem{LW} T.~D.~Lee and G.~C.~Wick, Phys.~Rev. D\textbf{2},
1033 (1970).

\bibitem{M} A.~Mostafazadeh, J.~Math.~Phys.~{\bf 43}, 3944 (2002).

\bibitem{BMW} C.~M.~Bender, P.~N.~Meisinger, and Q.~Wang,
J.~Phys.~A: Math.~Gen.~{\bf 36}, 1973 (2003).

\bibitem{BERRY} See, for example, M.~V.~Berry and C.~J.~Howls,
Proc.~Roy.~Soc.~Lond.~A {\bf 430}, 653 (1990).

\bibitem{BAN} A.~Banerjee, Preprint quant-ph/0502163 (2005). The
results in this preprint contain errors and cannot be used
directly to extract the semiclassical approximation to the $\cal
C$ operator.

\bibitem{B3} C.~M.~Bender, D.~C.~Brody, and H.~F.~Jones, Phys.
Rev.~Lett. \textbf{93}, 251601 (2004); Phys. Rev. D\textbf{70},
025001 (2004).

\bibitem{CCC} K.~Symanzik, Commun.~Math.~Phys. {\bf 45}, 79
(1975).

\bibitem{AAA} C.~M.~Bender, K.~A.~Milton, and V.~M.~Savage,
Phys.~Rev.~D {\bf 62}, 85001 (2000).

\bibitem{BBB} F.~Kleefeld, preprint hep-th/0506142 (2005).

\end{enumerate}
\end{document}